# Tunable narrowband Excitonic Optical Tamm States enabled by a metal-free all-organic structure

Miguel Castillo, Diogo Cunha, Carla Estevez-Varela, Daniel Miranda, Isabel Pastoriza-Santos, Sara Nuñez-Sanchez, Mikhail Vasilevskiy, Martin Lopez-Garcia*

**Abstract:** Optical Tamm States (OTS) are confined optical modes that can occur at the interface between two highly reflective structures. However, due to the strong reflectance required, their implementation with highly processable and metal-free materials has proven challenging. Herein, we develop the first structure supporting OTS based only on organic polymeric materials, demonstrating a photonic platform based on non-critical, widely available, and easily processable materials. The structures fabricated present large areas and consist of a narrowband multi-layered polymeric Distributed Bragg Reflector (DBR) followed by a thin film of J-aggregate molecular excitonic material that can act as a highly reflective surface within a narrowband range. We take advantage of the narrowband spectral response of the DBR and of the reflective molecular layer to tune the OTS band by varying the periodicity of the multilayer, opening the door for the fabrication of OTS structures based on lightweight integrable excitonic devices with cost-effective procedures.

## 1 Introduction

Light confinement into sub-wavelength scales is the core of the most relevant scientific achievements in photonics in recent years. The most common strategy to achieve effective field confinement is through the use of polaritonic states which can be generated via the coupling of light with a zoo of polaritonic particles such as excitons, electrons or phonons.[1] Surface plasmon polaritons in which photons and electrons are coupled at a metal surface[2] are by far the most studied example in nanophotonics given their wide range of applications, including nanoresolution, optical imaging and sensors.[3,4]

Optical Tamm States (OTS) have risen as an effective strategy for light confinement at the nanoscale without lateral nanostructuring requirements. In OTS, the light is confined at the interface of two highly reflective structures where the electric field decays exponentially with distance from this interface.[5] In their most simple but effective implementation, OTS have been shown to exist both between two Distributed Bragg Reflectors (DBR)[6] and between a DBR and metallic layer. Many applications of OTSs have been demonstrated, such as a directional emission by microlasers or single emitters[7,8] as well as angle-insensitive absorption structures.[9] OTS have also been used to achieve strong coupling with exciton-polariton states and inorganic quantum wells.[10] More recently, OTS-based thermal emitters in the infrared[11] and implemented in flexible metal films have been implemented.[12].

However, OTS, as well as many other polaritonic systems, have so far relied on materials with high refractive indices or negative values of the dielectric constant in the working spectral region. So far, these properties were only available by using contaminant or scarce materials such as semiconductors or noble metals[13] compassing complex processing methodologies. A fully soft-matter implementation of the OTS is much desirable because not only will it increase processability and sustainability, but also because it will open the possibility to implement polaritonic OTS devices in flexible or irregular-shaped surfaces such as buildings, furniture, or textiles. All-organic materials will increase processability and add great versatility to the structure since polymer films can usually be free-standing, flexible and their fabrication is less demanding than oxides enabling cleanroom-free fabrication. Still, the fabrication of OTS devices in all-soft matter platforms remains a challenging endeavour due to the relatively low refractive index of most polymeric materials.

In the most common implementation of an OTS device, a DBR is fabricated using inorganic materials of common use in optoelectronics, such as $TiO_2$, $SiO_2$ or $Si_3N_4$. The final DBR is coated by a thin metal layer acting as a top reflector. The reasons for these choices are twofold. First, large refractive index contrasts are often required to obtain wideband reflectance from the DBR that will allow the existence of an OTS at the metal-DBR interface. Secondly, these materials are suitable for fabrication over large areas with cleanroom processing techniques, allowing their integration with semiconductor technologies. However, these fabrication processes are also costly, time-consuming and environmentally damaging. Besides oxide dielectric mirrors do not allow the use of flexible substrates. Some alternatives to implementing metal-free OTS structures have been proposed where the metallic layer is substituted by doped graphene[14] or organic J-aggregate dyes.[15] However, these still rely on oxide thin film bottom mirrors meaning the aforementioned problems endure.

As a competitive alternative to inorganic DBR, liquid phase deposition by spin coating has been heavily studied. Often, these are hybrid organic/inorganic structures[16,17], but all-polymer systems have been demonstrated[18–24]. However, none of these has been used for the production of OTS. Usually, organic materials are not the first choice for light trapping structures because the low refractive index variability results in a low refractive index contrast. For that reason, nanoparticles have been used to dope organic layers of the system to increase this contrast.[25–27] This technique, however, presents its disadvantages such as a reduction in processability. Recently, all-organic photonic nanostructures with high contrast were demonstrated using J-aggregate doped polymers as near zero index materials increasing the refractive index contrast between polymers in well-defined spectral regions.[28,29] These J-aggregate doped polymers present metallic-like reflectance within a narrowband energy range, known as the Reststrahlen region.[30]. In this paper, we demonstrate that, when a J-aggregate doped polymer layer is combined with an all-organic DBR, it is possible to generate an OTS at the dye-DBR interface hence resulting in the possibility of a flexible and free-standing all-organic OTS structure. Firstly, we show how to design and fabricate a fully dielectric mirror by liquid phase spin coating deposition of polymer thin films. Secondly, we tune the structure's reflectance to the narrowband reflectance of the top J-aggregate-based layer. We show the feasibility of an all-organic OTS system and it should be noted that the properties described here can be extrapolated to similar structures with other polymer materials, especially to biopolymers which would open the door for OST biopolymer implementations for sustainable photonic devices.

## 2  Results and discussion

OTS occur at the interface between two strongly reflective structures, as demonstrated by the theoretical framework developed in the SI. In this work, the bottom mirror structure is an all-organic DBR composed of repeated unit cells of polystyrene (PS) and polyvinyl alcohol (PVA) layers, shown in **Figure 1a**.

The top mirror is a thin film of PVA doped with a J-aggregate supramolecular structure known as TDBC (see methods). The PS/PVA unit cell is repeated to provide strong enough reflectance to the bottom DBR to support OTS. The permittivity values of PS and PVA polymers are given in **Figure 1b** and are roughly constant along the VIS with values $\varepsilon_{PS} \approx 2.5$ and $\varepsilon_{PVA} \approx 2.2$, respectively. The absorption is negligible for both polymers, i.e. $Im(\varepsilon_{(PS/PVA)}) \approx 0$. Noteworthy, the permittivity contrast between PS and PVA is small ($\Delta\varepsilon \approx 0.3$) compared to non-organic alternatives (for example, $\Delta\varepsilon \approx 1.9$ for silicon dioxide and silicon nitride). Therefore, the final total number of layers of the bottom mirror is *100*, to achieve a strong band gap close with nearly 100% reflectance. Moreover, the low refractive index contrast creates a spectrally narrow photonic band gap when compared to the one observed in a DBR com-posed of oxide or semiconductor materials.[31] Therefore, the OTS structure will show an extraordinarily strong dependence on incident momenta as, to obtain a Tamm state, the photonic band gap of the DBR must match the optically metallic-like reflectance of the J-aggregate doped polymer film.

The top mirror is a thin film of TDBC doped PVA (TDBC-PVA). The optical properties of this polymer film can be modelled by a Lorentz oscillator with a narrow absorption band (≈35 meV) at a central vacuum frequency of $\omega_r$=2.1 eV (see **Figure 1b**). Here, we follow the previously reported fabrication protocol from ref.[32] to fabricate the TDBC-PVA layers (detailed description in methods). The real part of the permittivity of the films is highly dispersive, becoming negative between *2.*11 and *2.23* eV, see **Figure 1b**, yielding a strong reflectance on this narrow spectral range. In the following calculations, we considered the permittivity of the TDBC-PVA layers obtained from our previous work with samples produced by the same protocol.[29]

An OTS mode is supported by the two-mirror structure when the photonic bandgap of the DBR and the reflectance of the TDBC-PVA thin film match at a particular position in the energy and momentum space. The high reflectance band of the TDBC-PVA, between 2.11 and 2.23 eV, is independent of the incidence angle (or momentum) of incident light (red region in **Figure 1c**). However, the stop-band of the DBR blueshifts at higher momenta. In this paper, we have developed two OTS structures with two different PS/PVA unit cells to match the TDBC-PVA high reflectance band at two different momenta, one around 35° degrees and another around 45° degrees (see **Figure 1c**). The two designed OTS structures are formed by pairs of PS and PVA thin films with the thicknesses: $d_{PVA}$=105 nm and $d_{PS}$=99 nm (unit cell period *Δ=204 nm*) or $d_{PVA}$=101 nm and $d_{PS}$=95 nm (*Δ=196 nm*). **Figure 1c** illustrates the photonic stop-band in each case (green and yellow areas) as well as the spectral position for the high-reflectance band of the TDBC-PVA top mirror. Under these parameters, the photonic band gaps at normal incidence for the two structures are from *1.93* to *1.99* eV and *2.00* to *2.07* eV, respectively, both outside the high-reflectance region of the TDBC-PVA. However, the blueshift of the photonic band gap at higher momenta means that stop-band will eventually reach the optically metallic-like region of TDBC-PVA, at 45° and 35° as they were designed. The thickness of TDBC-PVA is fixed to $d_{TDBC\text{-}PVA}$=53 nm (see methods) after optimization via simulations for the best trade-off between the OTS field confinement and absorption upon illumination before reaching the interface between the two mirrors. For completeness, a detuned DBR was fabricated, with a photonic band gap at normal incidence from *2.40* to *2.47* eV, higher than the energies of the metallic-like region. Therefore the stop-band of the DBR will never overlap with the high-reflectance band of the TDBC-PVA mirror and hence not produce an OTS.

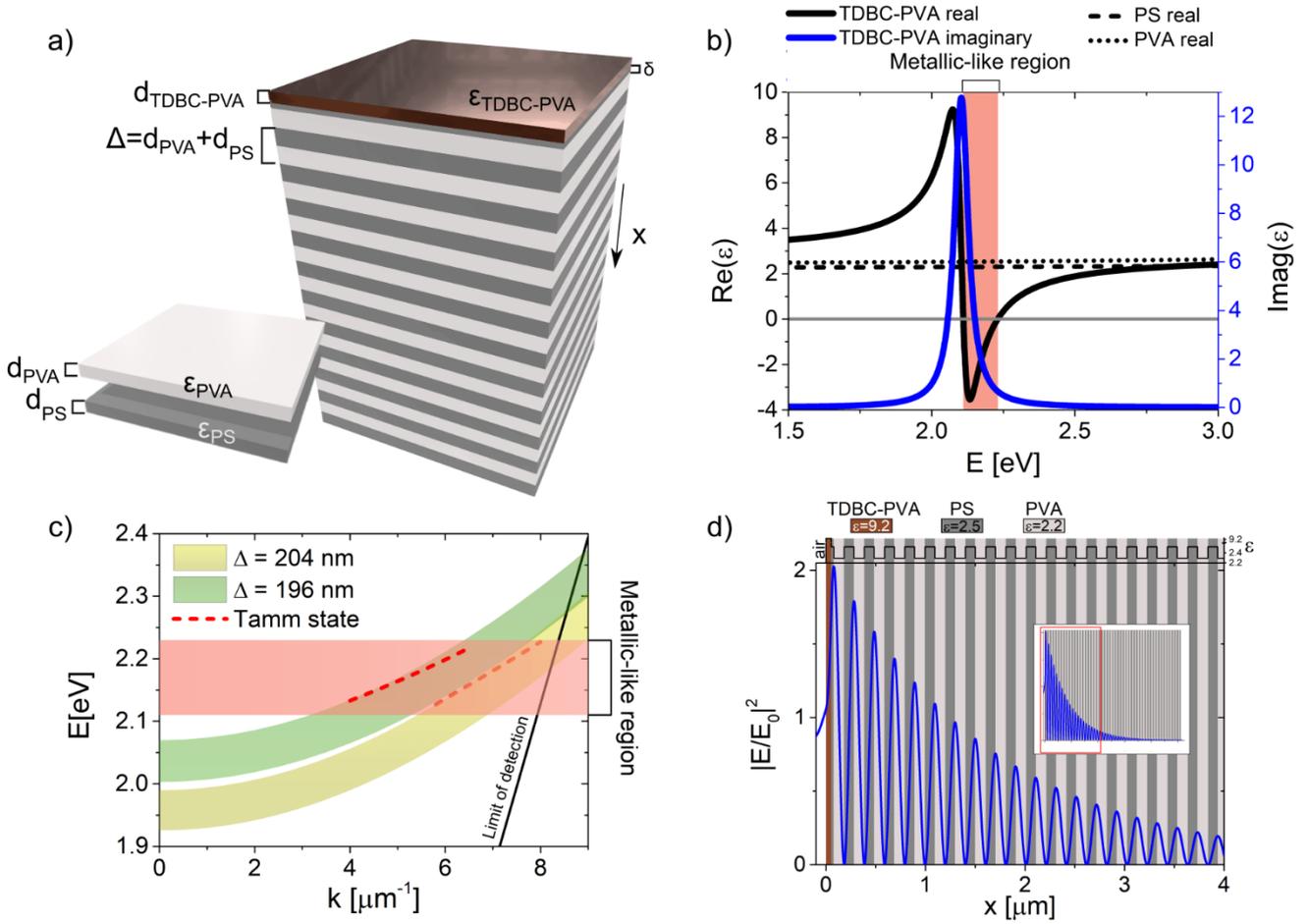

**Fig. 1: OTS-supported Structure. (a)** Schematic representation of the OTS structures consisting of two reflective sub-structures. The structure has in total 102 layers (100 layers for the DBR, a thin adhesion PS layer and a TDBC-PVA layer). **(b)** The complex dielectric constant of PVA (dash black line), PS (dotted black line) and TDBC-PVA films (blue and black continuous lines). The imaginary part of the dielectric constant for PVA and PS is zero in this wavelength range. **(c)** The photonic band gap of DBRs with $\Delta=204$ nm or 196 nm are indicated by yellow and green regions respectively. The high reflectance band of the TDBC-PVA film is represented by the red region. The OTS dispersion curves of the two structures are indicated by the dash red line. **(d)** Electric field intensity within the OTS structure for the sample $\Delta=204$ nm at $E=2.21$ eV and $k=7.91$ μm$^{-1}$ ($\vartheta=45°$). Light is incident from the air towards a glass substrate where the DBR was deposited.

The OTS dispersion curves (**Figure 1c**, red dotted lines) were estimated using a transfer matrix formalism (see supplementary information) for the two periods. Note that the limit of detection corresponds to the limit in the experimental set-up used to measure the samples that are fabricated in the next section. Our calculations show that the OTS field concentration **(Figure 1d)** is maximised when the last layer of the polymer DBR at the bottom mirror is the lowest index, in our case, the PVA ($\varepsilon_{PVA}<\varepsilon_{PS}$). These results are in agreement with other studies.[33] Note that a thin adhesion layer of PS between the TDBC-PVA film and the DBR was deposited with a thickness of $\delta\approx25$ nm. This was considered in the previous simulations, and it has a negligible effect on the photonic properties as is shown in **Figures S1a** and **S1b**. If we consider a real structure with finite mirrors, the OTS will be lossy due to radiative decay and its frequency will be a complex number, $\omega_r^*$, obeying the system of eqs. (S7) and (S8) with the corresponding transfer matrices and Fresnel coefficients. Since reflectivity measurements are done with propagating light possessing a real frequency, the reflectance minimum will occur at a frequency slightly deviated from $Re(\omega_r^*)$. If these calculations are performed without neglecting the exciton damping of TDBC-PVA, the reflectivity minimum also corresponds to a maximum of the absorbance. **Figure 1d** shows the exponentially decaying electric field at an OTS which concentrates at the interface between the two mirrors.

Considering the theoretical results (**Figure 1**), we fabricated an all-organic structure on glass consisting of PS/PVA alternating layers, with $\Delta=196$ nm, by liquid phase spin coating technique (see Methods). On top, a TDBC-PVA film is added, as a top mirror, by the same technique. In addition, each reflector was fabricated separately on a glass substrate for comparison. The *s*-polarised momentum-dependent reflectance of the two separate reflectors is shown in **Figure 2a** and the two combined in **Figure 2b.** The measurement is obtained through angle-resolved reflectance (see Methods). Single reflectance plots at normal and at an incident light $\vartheta=35°$ are also plotted in **Figures 2c-d**, respectively. Note that, the in-plane wave-vector at $\vartheta=35°$ incidence angle is given by $k=\omega/c.sin(\vartheta)$ and represented by a dash grey line in **Figure 2a-b**. The TDBC-PVA mirror shows a high reflectance band between 2.05 and 2.23 eV due to the characteristic metallic-like optical region of the material where $Re(\varepsilon)<0$ and due to the higher refractive index contrast caused by the dispersive Lorentzian resonance, seen on the left. We would like to remark this high reflectance band is independent of the in-plane momentum of the incoming light since there are no structural effects inherent to the optical properties of the layer. The DBR mirror shows a strong Bragg reflection band related to the photonic bandgap. The Bragg reflectance band blueshifts at higher momenta as was expected. Fabry-Perot oscillations are observed above and below the stop-band due to the finite thickness of the DBR (see Figs. 2c and d). Noteworthy, the Full-Width at half Maxima (FWHM) of the Bragg reflectance bands is around 0.06 eV, much narrower than other implementations, such as DBRs of silicon nitride and silicon dioxide with $FWHM=0.7$ eV.[15] This is due to the low refractive index contrast between PVA and PS organic materials. Despite this, DBR reflectance reaches values close to 100% illustrating the quality of the DBR structures fabricated, ensuring large enough photonic strength of the bottom mirror to generate an efficient OTS. We would like to remark that, due to the combination of a DBR with a narrow Bragg reflectance band and a top mirror with a narrow metallic-like reflectance, we are able to create OTS with narrow band response and with extreme precision of their angular response. The photonic bandgap and the TDBC-PVA metallic-like region cross for $3.7<k<8.0 \ \mu m^{-1}$ and $2.11<E<2.23$ eV, resulting in the OTS mode conditions being fulfilled for a limited range of angles of incidence (such as $\vartheta=35°$, shown by a dashed grey line).

**Figure 2b** shows the reflectance measurements and simulations in energy and momentum space for the two mirror structures combined. Both experimental and simulation results show that, for most $(k, E)$ values, reflectance shows the convolution of both responses described in Fig. 2a when the DBR stop-band and the TDBC-PVA reflectance do not match. But, a dip in reflectance is observed for $2.11<E<2.23$ eV which can be associated with the excitation of the OTS of the structure. The restriction of these features to the metallic-like spectral range (Reststrahlen band) of TDBC-PVA excludes other resonant phenomena, such as Fano-resonances. **Figure S1c** shows the same measurement for p-polarised light where, due to the narrower stop-band at higher momenta, the observation of the OTS is more challenging. **Figure S2** shows the same plot as **Figure 2b** but for the OTS structure with $\Delta=204$ nm. Because of the longer period, the overlap of the two reflective structures occurs at higher momenta, around $\vartheta=45°$.

To investigate further, we compare **Figure 2c**, normal incidence reflection, and **Figure 2d**, a cross-section represented by the grey dashed line ($\vartheta=35°$) where the photonic band gap is pushed towards the metallic-like region and hence an OTS state may exist. From here, there is a clear dip in the Bragg reflection where the electric field was found to be confined in **Figure 1d** at 2.21 eV. At this energy, the electric field is confined at the interface, reaching values of $|E|^2/|E_0|^2=2$. **Figure S3** shows that adding more layers to the structure does not affect the energy of the dip, though it displaces the Fabry-Perot oscillations, corroborating that this dip is due to an OTS and not a Fabry-Perot oscillation.

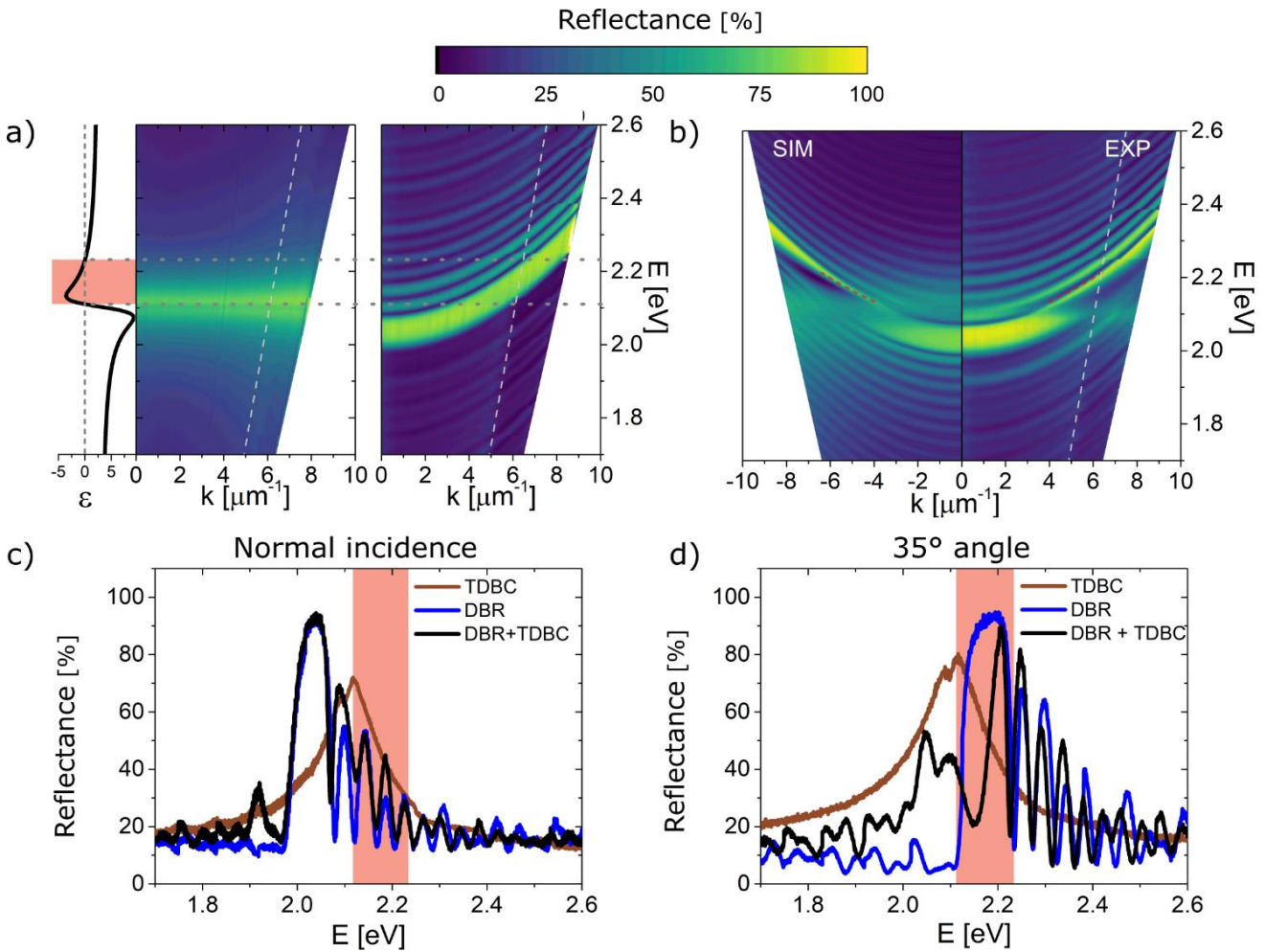

**Fig. 2: Reflectance measurements.** Experimentally measured angle-dependent absolute reflection spectrum (all s polarised) of the fabricated structure with $\Delta=196$ nm. **(a)** Thin TDBC-PVA film (left) and the DBR (right). To the left, the real permittivity is shown for reference. **(b)** The reflectance of OTS structure with $\Delta=196$ nm: simulation (SIM) and experimental (EXP). Red dashed lines show the solution of the OTS from equation S8 in SI. The wave-vector at $\theta=35°$ incidence angle is given by $k=\omega/c.\sin(\theta)$ and represented by a dash grey line. Cross-sections of these plots at normal incidence **(c)** and $\vartheta=35°$ **(d)**. The orange-shaded region represents the metallic-like area of TDBC-PVA where an OTS can occur.

PVA and PS are not absorbing polymers. Therefore, the field is concentrated at the interface between the TDBC-PVA mirror and the DBR mirror and can only be absorbed by the J-aggregate absorbing material, as it is calculated in **Figure S5**. Note that absorption is maximised at energies matching the OTS and not at the exciton energy ($E=2.10$ eV) where the imaginary component of the permittivity is at a maximum. Therefore, the absorption of the structure is dominated by the OTS mode and not the intrinsic absorption of the TDBC J-aggregate. Because J-aggregate dyes present strong photoluminescence (PL) response,[34] we can use PL to map experimentally the absorption enhancement induced by the Tamm states. **Figure 3a** shows the photoluminescence (PL) spectra of the samples fabricated: TDBC-PVA mirror and the OTS structures with $\Delta=196$ nm and $\Delta=204$ nm. The PL measurements were done using a Fluoro-MAX 3 spectrofluorometer (Horiba Scientific) with the samples placed at a $35°$ angle, providing high enough momentum to observe a Tamm state for the $\Delta=196$ nm sample, but too low that we can't observe it for $\Delta=204$ nm. Photoluminescence spectra of all samples were taken with an excitation energy of the light source of $E_{exc}=2.25$ eV. The PL spectra are similar between samples and peaked at $E=2.10$ eV, which is next to the exciton absorption band (represented in grey dashed line) due to the small stokes-shift of J-aggregates.[35] In the case of the OTS structure with $\Delta=196$ nm, which supports an OTS at 35º, a stronger PL peak intensity is obtained compared to the other OTS-containing structure, suggesting a potential enhancement of the PL intensity.

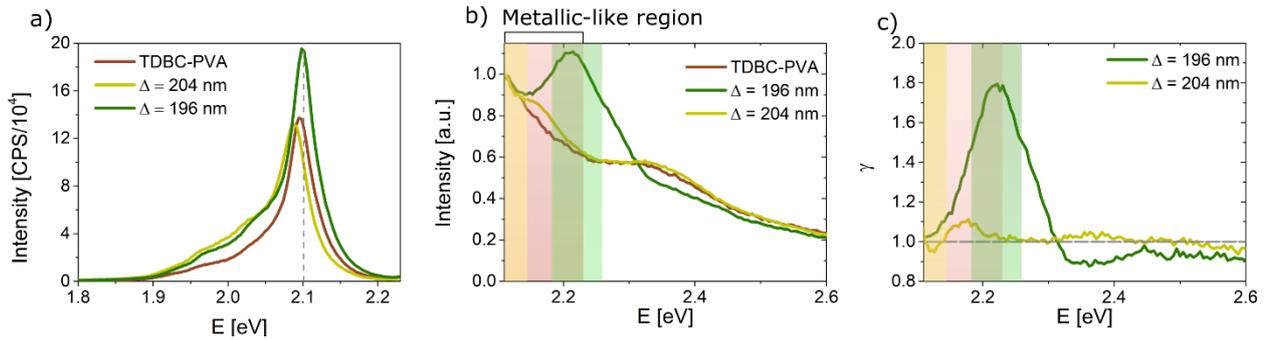

**Fig. 3: PL, PLE and enhanced absorbance of Tamm optical modes. a)** PL spectra under an excitation energy of $E_{exc}$=2.25 eV of a single TDBC-PVA layer and of the OTS structures. The dashed grey vertical line highlights the absorption maximum of the TDBC from the exciton band. **b)** PLE spectra averaged over 3 different points of a single TDBC-PVA layer and of the OTS structures. **c)** Absorption enhancement factor ($\gamma$) of the OTS structures caused by the presence of the DBR. Shaded areas correspond to the photonic band gaps of the DBRs. Note that all measurements (and consequently calculations of the photonic band gap) were performed at an incident light angle of 35 degrees. For clarity, each structure has a similar shadow colour as the DBR bandgap.

In addition, to corroborate if the absorbance of the TDBC-PVA layer is enhanced at the OST energies, we performed photoluminescence excitation (PLE) experiments, where the excitation light source was scanned in the range $1.97 < E_{exc} < 3.10$ eV. In **Figure 3b**, we compare the PLE of the three different samples at a 35 º incidence angle. Note that they have all been normalised to be unity at $E$=2.11 eV (start of the metallic-like region). The shaded regions represent the photonic stop-band of the DBR (dependent on the angle of incidence) and the optically metallic-like region of the TDBC-PVA top mirror (which is momentum independent, considering the exciton's kinetic energy to be negligible). As explained previously, the Tamm state will exist when there is an overlap between the stop-bands and the metallic-like region. Both samples show a partial overlap, however, only the sample with $\Delta$=196 nm shows this overlap in the higher energy region of the metallic-like area where a Tamm state can exist. Therefore, it becomes clear that, under these conditions, a Tamm state can only happen for the $\Delta$=196 nm. Note that, for the $\Delta$=204 nm sample we are showing only the range where it overlaps with the metallic-like region. However, this continues at energies below the start of the metallic-like region, which would not overlap and hence have a lighter shade; this is however not shown for clarity. As was expected, the PLE response of the OTS with $\Delta$=204 nm is similar to the TDBC-PVA reference slab since the sample doesn´t support any OTS mode at 35º. However, in the case of the OTS with $\Delta$=196 nm, a PLE peak is observed at 2.22 eV on top of a baseline defined by the TDBC-PVA absorbance. This absorption enhancement is caused by the light confinement on the absorbing layer due to the supported OTS mode. For clarity, we define a parameter where we normalise the PLE of the DBRs by the single layer, to calculate the absorption enhancement factor:

$$\gamma = PLE_{periodic\ structure} / PLE_{TDBC\text{-}PVA\ film}, \quad (2)$$

where the numerator represents the PLE of the complete structure (DBR and J-aggregate) and the denominator the PLE of the single TDBC-PVA layer. Therefore, when $\gamma > 1$, we can infer that we have absorption enhancement caused by the OTS multi-layer structure. In **Figure 3c** we can see this factor, where it becomes clear that PLE gets a boost of up to 80% in the region with a Tamm state, for the $\Delta$=196 nm sample. This factor quickly falls afterwards and for energies far from the gap, $\gamma \approx 1$ is obtained as there are negligible photonic effects. It can also be seen that $\Delta$=204 nm has $\gamma \approx 1$ for the entire spectrum due to the non-existence of significant photonic effects. Because the OTS coupling mode conditions require the spectral superposition of the optically metallic-like region of the top mirror and the bandgap of the bottom mirror, the structures are extremely sensitive to the incident angle. For this reason, an *8* nm difference between the structures is enough to show critical coupling for one structure and not the other. Therefore, we have developed metallic-free all-organic OTS structures whose absorption can be finely tuned by small changes in the periodicity of DBR. This could be achieved by local pressure or curvatures, making them fantastic candidates for free-standing coatings with an adaptable response.

## 3  Conclusion

In this paper, we showed, for the first time, that it is possible to obtain an OTS in all-organic structures. Our results show that the low refractive index contrast of all-organic Bragg reflectors can be a technological advantage to obtain, for example, narrowband OTS excitation. Moreover, the use of the optical metallic-like properties of a J-aggregate doped polymer layer (TDBC-PVA) adds a high sensitivity to in-plane momenta. Measurements of the DBR, the TDBC-PVA layer and the two combined layers confirm the presence of an OTS fulfilling the theoretical expectations. The photoluminescence excitation measurements corroborated that, at the OTS coupling conditions, the emission of the dye is increased suggesting the absorption is enhanced due to the excitation beam confinement. The strong modification of the emission properties due to the presence of the OTS opens the door for directional emission devices that could be explored in the future.

We would like to remark that the intrinsic narrowband and limited angular range of the all-organic OTS systems could be suitable for applications requiring coupling/decoupling directionality which could help avoid the lateral nanostructuring required in metal-based OTS devices. Moreover, the angle-dependent and narrow beam excitation range of the all-organic OTS system could be of interest for nanoscopy applications where angle and frequency selective illuminations are required and usually rely on bulk optics (such as in Total Internal Reflection fluorescence [36]). Traditional OTS have been proposed as cost-effective sensing platforms in the past, via monitoring changes in the reflectance upon changes in the medium (e.g. chemical components detection [37]). An all-organic and bio-compatible system, like the one presented here, offers added advantages such as in sensing applications by adding properties like flexibility or the possibility to introduce polymers reactive to specific chemical properties (e.g. pH) or local pressure in the structure.

Moreover, our results demonstrate that exotic optical modes and OTS-based devices can be fabricated with cost-effective and environmentally friendly materials, opening the door to more complex photonic structures, including resonant microcavities with enhanced emission and absorption performance. These developments will move photonics and polaritonic technologies to platforms based on non-critical raw materials which do not require mining, such as metals.

## 4  Methods

*Sample Preparation*. Every structure presented was fabricated on a coverslip (*170* μm thickness) by sequential deposition of polymeric layers by spin-coating. For the dye-doped metallic-like layers, we fabricated a matrix of J-aggregate (TDBC: 5,6-dichloro-2-[[5,6-dichloro-1-ethyl-3-(4-sulphobutyl)-benzimida-zol-2-ylidene]-propenyl]-1-ethyl-3-(4-sulphobutyl)-benzimida-zolium hydroxide, sodium salt, inner salt) with poly(vinyl alcohol) (PVA: Aldrich PVA Mw = *85000–124000*) for processability. We used a 3:1 mixture of *6.0%* wt. of PVA and *2%* wt. of TDBC following the well-established protocol.[15] The solution was then diluted 3:1 with water (1 part water) and stirred. Three DBRs (two tuned, *Δ=196* nm and $\Delta=204$ nm, and one detuned, $\Delta=161$ nm) were fabricated using the same polymers but different solution concentrations and RPMs to control the thickness of the layers and consequently the photonic bandgap. PS thin-films were deposited using a solution of polystyrene (Aldrich, $Mw\sim192$ kDa) diluted *2.9%* (both tuned) or 2.0% (detuned) wt. in toluene. PVA thin-films used a solution of poly(vinyl alcohol) (Aldrich PVA Mw = *33000*) diluted *6.0%* ($\Delta=204$ nm), *5.5%* ($\Delta=196$ nm) or *4.5%* ($\Delta=161$ nm) wt. in water. When fabricating multilayers by spin coating, the polymers used should be diluted by different solvents, such as polar and non-polar. This ensures minimal damage to the underlying layer enabling processability and flexibility.[18] TDBC-PVA is deposited at *4500* rpm to yield a *53* nm thick layer. PS was deposited at *5250* ($\Delta=204$ nm), *6000* ($\Delta=196$ nm) or *3000* ($\Delta=161$ nm) rpm, yielding *99*, *94* and *78* nm respectively, and PVA at *5750* ($\Delta=204$ nm), *5000* ($\Delta=196$nm) or *4000* ($\Delta=161$ nm) rpm, yielding *105*, *100* and *83* nm respectively. To fabricate the *25* nm layer of PS separating PVA and TDBC-PVA, a *1%* wt. PS solution at *6000* rpm was used.

*Reflectance Measurements*. A high-magnification optical microscope coupled to a Fourier image spectroscopy setup[36] was used to characterize the samples fabricated. The system uses a tungsten-halogen white light lamp covering the UV-VIS-NIR spectral range. The angle-dependent reflection measurements were probed and collected using a high numerical aperture (NA) lens (Nikon Plan Fluor 40x, NA=0.75 OFN25 DIC M/N2). The size of the spot on the sample was *40 μm*. These reflection measurements were collected using a spectrograph (Princeton Instruments, Acton SpectraPro SP-2150) and a CCD camera (QImaging Retiga R6 USB3.0 Color). Reflection measurements were normalized against the reflection of an optically thick silver mirror.

*Numerical calculations.* The numerical calculations of the reflectance, absorptance and electric field shown were done using a Transfer Matrix Method (TMM)[31] written in-house with Python. The photonic band gaps were calculated using a program based on the Bloch theorem. The OTS was calculated from an analytical formula described in the supplementary.[37]

# Author's information

- Miguel Castillo, Daniel Miranda, Martin Lopez*

  Natural and Artificial Photonic Structures Group. International Iberian Nanotechnology Laboratory, Braga 4715-330, Portugal.

  *Email: martin.lopez@inl.int

- Miguel Castillo

  Faculty of Physics/Faculty of Optics and Optometry, Campus Vida s/n, University of Santiago de Compostela, E-15782 Santiago de Compostela, Galicia, Spain

- Diogo Cunha and Mikhail Vasilevskiy

  Centro de Física das universidades do Minho e do Porto, Laboratório de Física para Materiais e Tecnologias Emergentes (LaPMET), Universidade do Minho, Braga 4710-057, Portugal

- Carla Estevez-Varela, Isabel Pastoriza-Santos, Sara Nuñez-Sanchez

  CINBIO, Universidade de Vigo, Departamento Química Fisica, 36310 Vigo, Spain

  *Email: S.Nunez-Sanchez@uvigo.es

- Mikhail Vasilevskiy

  Departamento de Física, Universidade do Minho, Braga 4710-057, Portugal

  Theory of Quantum Materials Group. International Iberian Nanotechnology Laboratory, Braga 4715-330 Portugal

# Supporting information

Data is made available upon request to MLG.

## Acknowledgements

The work by M.C.A, M.L-G and D.M was supported by the "Towards Biomimetic Photosynthetic Photonics" project (POCI-01-0145-FEDER- 031739) co-funded by FCT and COMPETE2020.C. S.N.-S., C.E.-V. and I.P.-S. acknowledge financial support from MCIN/AEI/10.13039/501100011033 (Grant No. PID2019-108954RB-I00 and PRE2020-096163) and the Xunta de Galicia/FEDER (grant GRC ED431C2020/09). D.C and M.V. acknowledge funding from the Portuguese Foundation for Science and Technology (FCT) in the framework of the Strategic Financing UID/FIS/04650/2020.

## Conflict of interest

The authors declare no conflicts of interest.

# Supporting Information

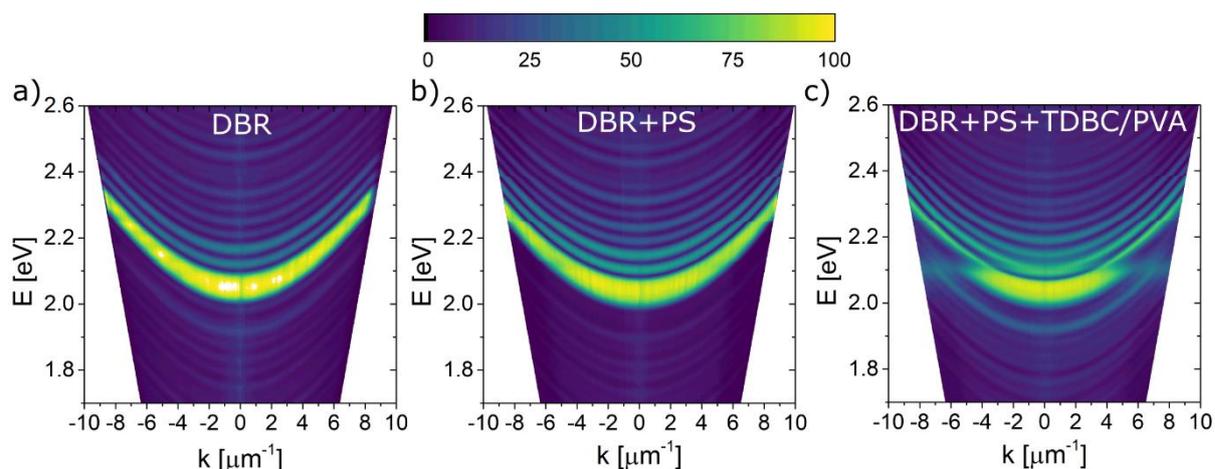

**Figure S1: p polarised reflection measurements**. Angle-dependent reflection spectrum (p polarised) of the DBR (a), DBR with thin PS layer (b) and DBR with thin PS layer and TDBC-PVA (c), represented in figure 1b, with $\Delta$=196 nm of the main manuscript.

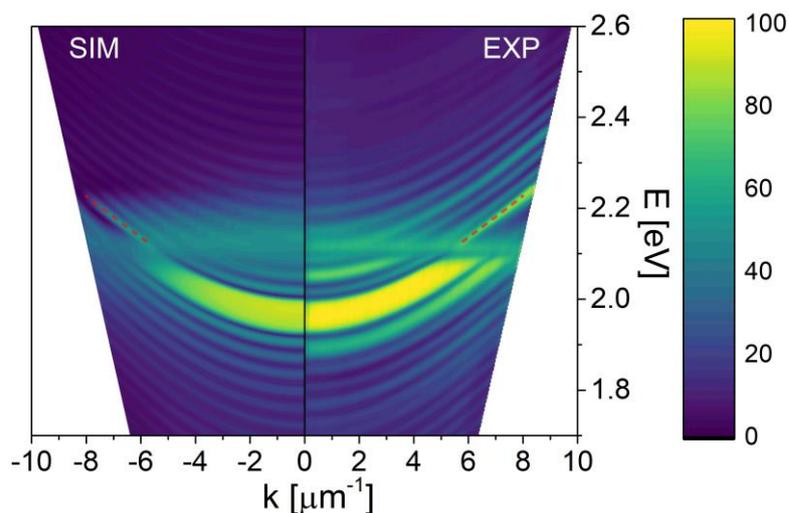

**Figure S2: Reflection of *204* nm period sample.** Simulated and measured angle-resolved reflection spectrum (s polarised) of the $\Delta$=204 nm DBR sample. Red dashed lines represent the solution from equation S8.

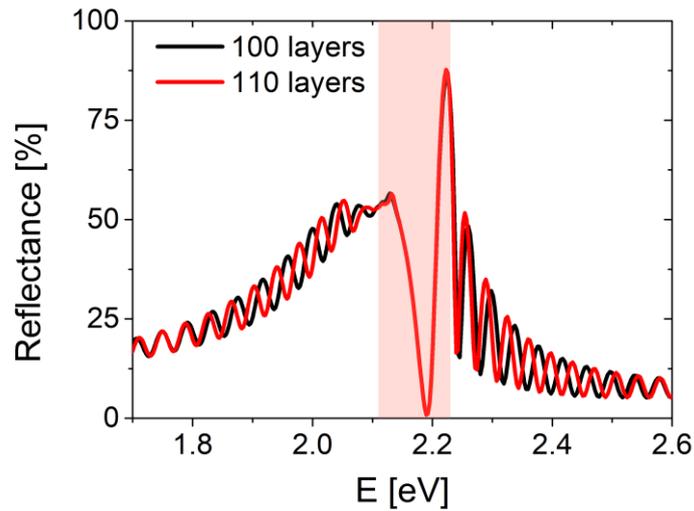

**Fig S3. Fabry-Perot oscillations.** S polarised reflection at *35°* of two DBRs with *Δ=196* nm with a different number of layers. The shaded region represents the optically metallic-like area of TDBC-PVA where a Tamm state can occur.

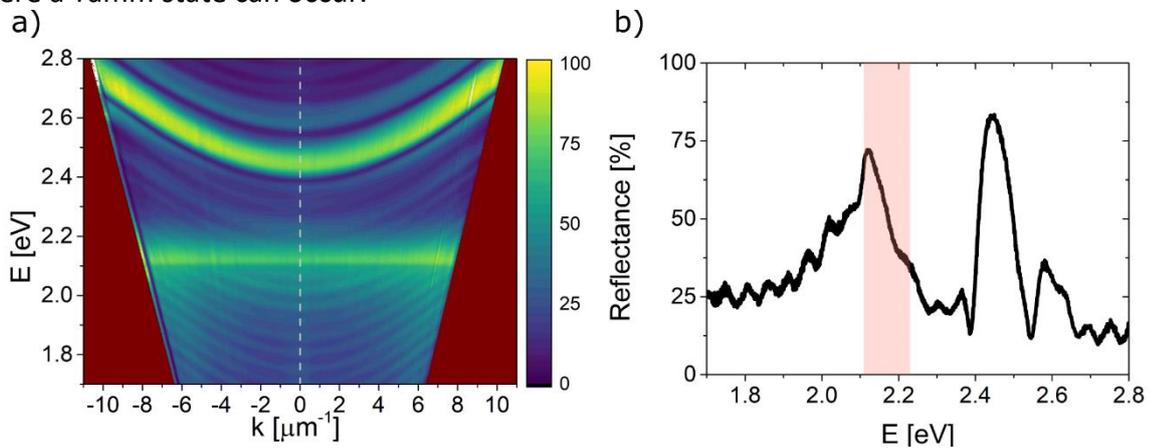

**Fig S4. Detuned photonic bandgap.** Angle dependent (s polarised) (a) and normal incidence (b) reflection spectrum of a structure similar to that represented in figure 1 of the main manuscript, but with a detuned photonic bandgap and metal-like optical properties of TDBC-PVA. This detuned DBR has thicknesses $d_{PVA}$=*83* nm and $d_{PS}$=*78* nm (producing a photonic bandgap at *2.48* eV at normal incidence).

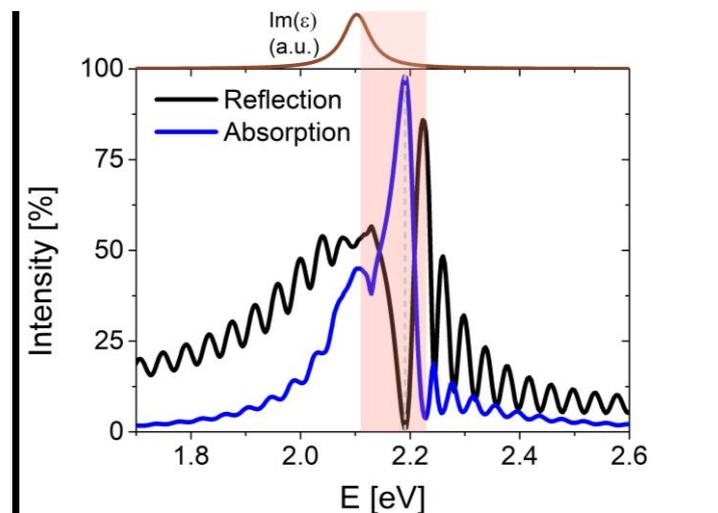

**Figure S5. Tamm state enhancing absorption.** Simulated s polarised reflection and absorption at $\theta = 35°$ of the structure in figure 1 with *Δ=196* nm. The shaded region represents the optically metallic-like area of TDBC-PVA and the vertical grey dashed line represents the minimum in reflection.

**Theoretical calculation of the Optical Tamm state**

To calculate the reflectance of these structures, we make use of the transfer matrix formalism. It allows us to evaluate the Fresnel coefficients from the matrix $T_{ij}$ that transforms in-plane components of the incident electromagnetic field into the transmitted ones. For the *s*-polarization, the Fresnel coefficients are written as:

$$\hat{t}_s = \frac{2}{T_{11}^{-1} + \frac{ck_{3x}}{\omega}T_{12}^{-1} + \frac{\omega}{ck_{1x}}T_{21}^{-1} + \frac{k_{3x}}{k_{1x}}T_{22}^{-1}} \quad (S1)$$

$$\hat{r}_s = \left(T_{11}^{-1} + \frac{ck_{3x}}{\omega}T_{12}^{-1}\right)\hat{t}_1 - 1 \quad (S2)$$

where $k_{1(3)x}$ is the wave vector of the incident (transmitted) wave and $T_{ij}^{-1}$ are the matrix elements of the inverse transfer matrix of the whole structure,

$$T^{-1} = T_{TDBC}^{-1}.T_{DBR}^{-1}, \quad (S4)$$

with $T_{DBR}^{-1} = (T_{PVA}^{-1}.T_{Ps}^{-1})^N$, where $N$ is the number of pairs of layers (i.e. periods) constituting the Bragg reflector and the individual matrices, for a medium A, are written as

$$\hat{T}_A = \begin{pmatrix} \cos(k_{Ax}d_A) & i\frac{\omega}{ck_{Ax}}\sin(k_{Ax}d_A) \\ i\frac{ck_{Ax}}{\omega}\sin(k_{Ax}d_A) & \cos(k_{Ax}d_A) \end{pmatrix}. \quad (S5)$$

Here $d_A$ is the thickness of the layer and $k_{Ax} = (\varepsilon_A(\omega/c)^2 - k^2)^{1/2}$ with $k$ being the transversal wave-vector and c the light velocity.

With the TDBC-PVA layer included in the structure, a minimum in the reflectance spectra appears within the optical metal-like region of TDBC-PVA, which is a fingerprint of the Tamm state. The OTS is an eigenstate of the whole structure and it can be shown that its frequency, for a given $k$, can be found by equating the denominator of Eq. (S1) to zero.[37] This condition can be cast in the following form:[33]

$$\hat{r}_1\hat{r}_2e^{2ik\delta} = |\hat{r}_1|e^{i\varphi_1}|\hat{r}_2|e^{i\varphi_2}e^{2ik\delta} = 1 \quad (S5)$$

where $\hat{r}_1$ ($\hat{r}_2$) corresponds to the Fresnel reflection coefficient of the heterostructure 1(2) alone (in vacuum) and $\delta$ is the thickness of the buffer layer. For simplicity, we will consider that the bottom mirror is the heterostructure 2 (dictated by the reflection coefficient $\hat{r}_2$) and the heterostructure 1 is the TDBC-PVA layer alone. It means that, under these assumptions, we can consider the limit of $\delta \to 0$ and Eq. (S5) reduces to:

$$\hat{r}_1\hat{r}_2 = |\hat{r}_1|e^{i\varphi_1}|\hat{r}_2|e^{i\varphi_2} = 1 \quad (S6)$$

From Eq. (S6) we have:

$$\begin{cases} |\hat{r}_1||\hat{r}_2| = 1 & (S7) \\ \varphi_1 + \varphi_2 = 2\pi m & (S8) \end{cases}$$

where $m$ is an integer and $\varphi_{(1(2))}$ is the argument of the complex number $\hat{r}_{1(2)}$. Eq. (S8) is known as the phase-matching condition and it determines an implicit dispersion relation of the OTS. For semi-infinite DBR and TDBC-PVA, with the imaginary parts of the permittivity neglected, Eq. (S7) is obeyed automatically and the phase-matching condition (S8) yields a real eigen-frequency. To calculate it, we need to take the limit $N\to\infty$ for evaluating the DBR transfer matrix (which is facilitated by using the Chebyshev formula) and, for the TDBC-PVA layer, take the reflection Fresnel coefficient as

$$\hat{r}_1 = \frac{k_{1x} - k_{2x}}{k_{1x} + k_{2x}}, \quad (S9)$$

where $k_{1x}$ and $k_{2x}$ are the wave-vector's normal components in air and TDBC-PVA, respectively.